\documentclass[12pt]{article}
\usepackage{a4wide,graphicx,subfig}
\usepackage[hyperindex]{hyperref}
\usepackage{amsmath,amsthm,amsfonts}
\usepackage{natbib}
\usepackage{pgfplots}
\usepgfplotslibrary{fillbetween}

\hypersetup{
colorlinks=true,
citecolor=green,
linkcolor=blue,
urlcolor=magenta
}

\newtheorem{theorem}{Theorem}[section]
\newtheorem{remark}[theorem]{Remark}

\newcommand{\ExT}[3]{\mathbb{E}_{#1}^{#2}\!\left[\,#3\,\right]}

\newcommand{\ind}[1]{1_{\{#1\}}}

\title{Funding Adjustments in Equity Linear Products}
\author{
Stefania Gabrielli\thanks{Be Consulting, Piazza Affari 2, 20123 Milano, Italy. Email: {\tt s.gabrielli@be-tse.it}.}
\ \ \
Andrea Pallavicini\thanks{Department of Mathematics, Imperial College, London SW7 2AZ, UK and Banca IMI, Largo Mattioli 3, 20121 Milano, Italy. Email: {\tt a.pallavicini@imperial.ac.uk}.}
\ \ \
Stefano Scoleri\thanks{Be Consulting, Piazza Affari 2, 20123 Milano, Italy. Email: {\tt s.scoleri@be-tse.it}, corresponding author.}
}

\date{
\small First Version: January 11, 2019. This Version: \today
}

\begin{document}

\maketitle

\begin{abstract}

Valuation adjustments are nowadays a common practice to include credit and liquidity effects in option pricing. Funding costs arising from collateral procedures, hedging strategies and taxes are added to option prices to take into account the production cost of financial contracts so that a profitability analysis can be reliably assessed. In particular, when dealing with linear products, we need a precise evaluation of such contributions since bid-ask spreads may be very tight. In this paper we start from a general pricing framework inclusive of valuation adjustments to derive simple evaluation formulae for the relevant case of total return equity swaps when stock lending and borrowing is adopted as hedging strategy.

\end{abstract}

\bigskip

\noindent {\bf AMS classification codes:} 91G20.\\
\noindent {\bf Keywords:} Funding, Valuation Adjustments, FVA, Collateral, Equity Swap, Total Return Swap, Stock Lending, Dividend Tax, Tobin Tax.

\newpage
{\small \tableofcontents}
\vfill
{\footnotesize \noindent The opinions here expressed  are solely those of the authors and do not represent in any way those of their employers.}
\newpage

\pagestyle{myheadings} \markboth{}{{\footnotesize  S. Gabrielli, A. Pallavicini, S. Scoleri, Funding Adjustments in Equity Linear Products}}

\section{Introduction}
\label{sec:introduction}
After the Financial crisis started in 2008, it was realized that the collateralization and funding mechanisms of OTC derivatives typically have a sizeable effect on their valuation. It is now customary to adjust the risk-free value of a derivative contract by adding some quantities, collectively known as XVAs, in order to deal with such effects and to charge the counterparty with the corresponding costs. Recent literature on the subject can be found e.g. in \cite{Brigo2019}, \cite{Bielecki2018}, \cite{Crepey2015}, \cite{Capponi2018}, \cite{Burgard2013}.

In the present work, we consider the problem of pricing linear equity products including all funding sources, both on the derivative side and on its hedge. In particular, we focus on Total Return Swaps (TRS), where the total return on an underlying asset is exchanged against floating cash-flows (LIBOR plus spread). A bank usually hedges a TRS with an opposite position in the underlying asset, so as to replicate the derivative cash-flows. In particular, stock lending and borrowing are commonly used as hedging strategies, even though direct purchase of the stock (buy-and-hold) is possible in principle. All the financing costs generated by the chosen strategy should be carefully taken into account and included into the TRS spread. These costs typically come from funding the collateral accounts, from taxes paid on dividends and other forms of taxation such as the so-called Tobin tax\footnote{In the European Union, a Tobin-style tax on financial transactions (FTT) has been first proposed in 2011. Currently, it is adopted by 10 Member States. See \url{https://ec.europa.eu/taxation_customs}.}.  

Notice that, even when the TRS is fully collateralized, the hedge is not, since a haircut is applied to repo transactions. As a result, funding unsecured is always needed in a certain amount, giving rise to an FVA contribution. These effects are reflected in the disounting curves used for pricing, as we will detail in the following sections.

Another peculiarity of funding equity TRS contracts is the role played by taxes. We can identify four different tax contributions, arising from the derivative and hedging mechanics: first of all, the reduction of dividend flows due to taxes comes with different amounts depending on the party receiving the dividends (the stock borrower in the repo market, the investor in the stock market, the equity receiver in the TRS market). We show how the TRS price is impacted by these tax asymmetries. Secondly, the EU FTT implemetation of the Tobin tax requires that a percentage of the stock market value, ranging from 10 to 20 basis points in most markets, is paid each time a certain amount of shares is bought, e.g. for hedging purposes. In some cases, the Tobin tax contribution cannot be neglected since it is of the same order of the TRS spread.

The paper is organized as follows: in section \ref{sec:forward} we set up the theoretical pricing framework, extending \cite{Duffie2001} approach to the presence of collateral agreements in different currencies, and apply it to the computation of equity forward prices in general situations. In section \ref{sec:trs} we apply our pricing framework to the case of Total Return Swaps and compute the TRS spread inclusive of all funding costs when different hedging strategies are adopted. In section \ref{sec:numeric} we support the results obtained in the previous section with some numerical examples and in section \ref{sec:conclusion} we draw our conclusions. Similar topics are covered e.g. in \cite{Fries2014} and in \cite{Lou2018}.     

\section{Forward Prices}
\label{sec:forward}

We consider an asset traded on the market. We term respectively $S_t$ and $D_t$ the price and cumulative dividend processes. If the asset requires a margining procedure, we term $C_t$ the price process of the margin account. Trading on the market requires a funding bank account. We term $B_t$ its price process. Here we consider symmetric rates, i.e. lending and borrowing are assumed to be made at the same rate. Furthermore, CVA and DVA are not considered in the present analysis. These assumptions can be relaxed, see e.g. \cite{Brigo2018}. 

We can normalize asset prices w.r.t. the bank account by defining the deflated price and dividend processes. In general, we assume that all deflated processes are expressed in the same currency of the bank account, on the other hand dividends and periodic margin payments may be expressed in a different currency. We can define (see \cite{Duffie2001} and \cite{MoreniPallavicini2015}) the deflated price and cumulative dividend processes as given by
\begin{equation}
{\bar S}_t := \frac{\chi^{\rm f}_t}{B_t} S^{\rm f}_t
\end{equation}%
and
\begin{equation}
{\bar D}_t := \int_0^t \frac{\chi^{\rm f}_u}{B_u} \,d\pi^{\rm f}_u + \int_0^t \frac{\chi^{\rm g}_u}{B_u} \left( dC^{\rm g}_u - c^{\rm g}_u C^{\rm g}_u \,du + d\langle \log \chi^{\rm g}, C^{\rm g} \rangle_u \right)
\end{equation}%
where $\rm f$ and $\rm g$ refers respectively to the dividend currency and to the margin account currency with $\chi^{\rm f}_t$ and $\chi^{\rm g}_t$ the corresponding FX spot prices which convert the cash flow in the bank account currency. The terms in the integral have the following meaning: (i) contractual coupons, (ii) collateral posts due to the margining procedure, (iii) additional fees to match the margin account accrual rate, (iv) wrong-way risk of funding a margin account in a currency different from the bank account one. We can define also a deflated margin account price process as given by
\begin{equation}%
{\bar C}_t := \frac{\chi^{\rm g}_t}{B_t} C^{\rm g}_t
\end{equation}%

The profit-and-loss generated by a buy-and-hold strategy is defined as the gain process $G_t$. The deflated gain process can be written as
\begin{equation}
{\bar G}_t := {\bar S}_t + {\bar D}_t - {\bar C}_t
\end{equation}%

In order to avoid classical arbitrages among all the admissible trading strategies we require the existence of a risk-neutral measure $\mathbb Q$ equivalent to the physical one such that ${\bar G}_t$ is a martingale under such measure. A direct consequence of the martingale condition is that we can calculate the growth rate of the asset under the risk-neutral measure. Indeed, we have
\begin{eqnarray}
B_t \,d{\bar G}_t
&=& B_t \left( d{\bar S}_t + d{\bar D}_t - d{\bar C}_t \right) \\\nonumber
&=& - \chi^{\rm f}_t r_t S^{\rm f}_t \,dt + \chi^{\rm f}_t dS^{\rm f}_t + S^{\rm f}_t d\chi^{\rm f}_t + d\langle \chi^{\rm f},S^{\rm f} \rangle_t + \chi^{\rm f}_t \,d\pi^{\rm f}_t + \chi^{\rm g}_t \left(r_t-c^{\rm g}_t\right) C^{\rm g}_t \,dt - C^{\rm g}_t d\chi^{\rm g}_t
\end{eqnarray}%
Then, we assume under the risk-neutral measure in the bank-account currency the following general dynamics for the asset and the FX spot prices
\begin{equation}
dS^{\rm f}_t = S^{\rm f}_t \mu^{\rm f}_t \,dt + dM^{\rm f}_t
\;,\quad
d\chi^{\rm f}_t = \chi^{\rm f}_t \nu^{\rm f}_t \,dt + dN^{\rm f}_t
\;,\quad
d\chi^{\rm g}_t = \chi^{\rm g}_t \nu^{\rm g}_t \,dt + dN^{\rm g}_t
\end{equation}
where $M_t$ and $N_t$ are two risk-neutral martingales. Thus, we can write
\begin{equation}
B_t \,d{\bar G}_t
= \chi^{\rm f}_t \,d\pi^{\rm f}_t + \chi^{\rm f}_t S^{\rm f}_t \left( \mu^{\rm f}_t + \nu^{\rm f}_t - r_t \right) dt + d\langle \chi^{\rm f},S^{\rm f} \rangle_t - \chi^{\rm g}_t C^{\rm g}_t \left( c^{\rm g}_t + \nu^{\rm g}_t - r_t \right) dt + \ldots
\end{equation}%
where the dots on the right-hand side represent the martingale part. Hence, by taking risk-neutral expectations of both sides, and using the martingale property, we get
\begin{equation}
\mu^{{\rm f},{\mathbb Q}}_t := \mu^{\rm f}_t
= r_t - \nu^{\rm f}_t - \theta_t - \frac{ \partial_t \pi^{\rm f}_t }{ S^{\rm f}_t } + \frac{ \chi^{\rm g}_t C^{\rm g}_t }{ \chi^{\rm f}_t S^{\rm f}_t } \left( c^{\rm g}_t + \nu^{\rm g}_t - r_t \right)
\end{equation}%
where we define the It\^o correction
\begin{equation}
\theta^{\rm f}_t \,dt := \frac{d\langle \chi^{\rm f},S^{\rm f} \rangle_t}{\chi^{\rm f}_tS^{\rm f}_t}
\end{equation}
which represents the drift correction of an asset in currency $\rm f$ when observed under the risk-neutral measure in the bank-account currency. If we change measure to the risk-neutral measure in currency $\rm f$ we finally obtain
\begin{equation}
\mu^{{\rm f},{\mathbb Q}^f}_t = r_t - \nu^{\rm f}_t - \frac{ \partial_t \pi^{\rm f}_t }{ S^{\rm f}_t } + \frac{ \chi^{\rm g}_t C^{\rm g}_t }{ \chi^{\rm f}_t S^{\rm f}_t } \left( c^{\rm g}_t + \nu^{\rm g}_t - r_t \right)
\end{equation}%

We can now calculate forward prices $F^{\rm f}_t(T)$ by substituting the expression for $\mu^{{\rm f},{\mathbb Q}^f}_t$ into the asset price dynamics under the risk-neutral measure in currency $\rm f$.
\begin{equation}
\partial_T F^{\rm f}_t(T) = \ExT{t}{{\mathbb Q}^f}{ \left( r_T - \nu^{\rm f}_T \right) S^{\rm f}_T - \partial_T \pi^{\rm f}_T + \frac{ \chi^{\rm g}_T}{ \chi^{\rm f}_T } C^{\rm g}_T \left( c^{\rm g}_T + \nu^{\rm g}_T - r_T \right) }
\end{equation}
In case of deterministic interest-rates we get
\begin{equation}
\partial_T F^{\rm f}_t(T) = \left( r_T - \nu^{\rm f}_T \right) F^{\rm f}_t(T) - \partial_T \,\ExT{t}{{\mathbb Q}^f}{ \pi^{\rm f}_T } + \ExT{t}{{\mathbb Q}^f}{ \frac{ \chi^{\rm g}_T }{ \chi^{\rm f}_T } C^{\rm g}_T } \left( c^{\rm g}_T + \nu^{\rm g}_T - r_T \right)
\end{equation}

We assume that the contractual dividends are constituted by absolute dividends $q^{\rm h}_k$ (possibly expressed in currency $\rm h$) plus a proportional repo fee $\ell^{\rm f}_t$, while margins are always proportional.
\begin{equation}
\pi^{\rm f}_t = \int_0^t \ell^{\rm f}_u S^{\rm f}_u \,du + \sum_k \frac{ \chi^{\rm h}_{t_k} }{ \chi^{\rm f}_{t_k} } q^{\rm h}_k \ind{t>t_k}
\;,\quad
C^{\rm g}_t = (1+\alpha_t) \frac{ \chi^{\rm f}_t }{ \chi^{\rm g}_t } S^{\rm f}_t
\end{equation}%

\begin{remark}
The assumption of proportional collateralization can be questionable since it depends on the funding costs of the investor. Alternatively we can introduce a security price process subject to perfect collateralization (i.e. haircut $\alpha_t$ equal to zero), and use it to evaluate the collateral in the original problem. See \cite{Fries2014}.
\end{remark}

We can proceed by defining the net $q$ and gross $Q$ forward dividend price in foreign currency $\rm f$ as given by
\begin{equation}
q^{\rm f}_t(t_k) := \ExT{t}{{\mathbb Q}^f}{ \frac{ \chi^{\rm h}_{t_k} }{ \chi^{\rm f}_{t_k} } q^{\rm h}_k }
\;,\quad
Q^{\rm f}_t(t_k) := \frac{1}{1-\rho} \,q^{\rm f}_t(t_k)
\end{equation}
where $\rho$ is the dividend tax (or other contractual reductions). Then, we substitute the above formula in the forward expression to get
\begin{equation}
\partial_T F^{\rm f}_t(T) = \left[ -\alpha_T (r_T - \nu^{\rm f}_T) + (1+\alpha_T) \left( c^{\rm g}_T + \nu^{\rm g}_T - \nu^{\rm f}_T \right) - \ell^{\rm f}_T \right] F^{\rm f}_t(T) - \sum_k q^{\rm f}_t(t_k) \delta(T-t_k)
\end{equation}
We can define the effective funding and collateral curves in currency $\rm f$ as given by
\begin{equation}
r^{\rm f}_t := r_t - \nu^{\rm f}_t
\;,\quad
c^{\rm f}_t := c^{\rm g}_t + \nu^{\rm g}_t - \nu^{\rm f}_t
\end{equation}%
as long as the repo-adjusted blended discounting curve in currency $\rm f$
\begin{equation}\label{eq:reporate}
z^{\rm f}_t := -\alpha_t r^{\rm f}_t + (1+\alpha_t) c^{\rm f}_t - \ell^{\rm f}_t
\end{equation}%
to write the forward price ODE
\begin{equation}
\partial_T F^{\rm f}_t(T) = z^{\rm f}_T F^{\rm f}_t(T) - \sum_k q^{\rm f}_t(t_k) \delta(T-t_k)
\end{equation}%
which can be solved to obtain
\begin{equation}
F^{\rm f}_t(T) = \frac{S^{\rm f}_t}{P_t(T;z^{\rm f})} - \sum_k \frac{P_t(t_k;z^{\rm f})}{P_t(T;z^{\rm f})} q^{\rm f}_t(t_k) \ind{t_k<T}
\end{equation}%
where zero-coupon bond prices with yield $z_t$ are defined as
\begin{equation}
P_t(T;z^{\rm f}) := \exp\left\{ -\int_t^T z^{\rm f}_u \,du \right\}
\end{equation}%

\subsection{Buy-and-Hold Strategy}
Consider the problem of hedging a forward contract in which at maturity you have to deliver the asset in change of cash. The simplest strategy is buying at contract inception the asset and holding it up to maturity. In a buy-and-hold (BH) strategy we use the bank account $B_t$ for any funding needs. We do not have any margin or repo fee to pay (or receive), but we have to pay taxes on dividends, namely we have
\begin{equation}
z^{\rm f}_t \doteq r^{\rm f}_t
\;,\quad
q^{\rm h}_k \doteq \left(1 - \rho_{\rm I}\right) Q^{\rm h}_k
\end{equation}%
where $\rho_{\rm I}$ is the dividend taxation the investor is subjected to, and $Q^{\rm h}_k$ is the gross dividend. Thus, we obtain
\begin{equation}
F^{\rm f}_t(T) = \frac{S^{\rm f}_t}{P_t(T;r^{\rm f})} - \sum_k \frac{P_t(t_k;r^{\rm f})}{P_t(T;r^{\rm f})} \left(1 - \rho_{\rm I}\right) Q^{\rm f}_t(t_k) \ind{t_k<T}
\end{equation}%

\subsection{Stock Lending Strategy}
Consider, as in the previous example, the problem of hedging a forward contract in which at maturity you have to deliver the asset in change of cash. This time at inception you buy the asset and lend it to a third party (the borrower) up to maturity. In a stock lending (SL) strategy we need to pay the interests on cash collateral while receiving the repo fees. Only a fraction of the dividends are paid back to the lender. We have
\begin{equation}
q^{\rm h}_k \doteq \left(1 - \rho_{\rm B}\right) Q^{\rm h}_k
\end{equation}%
where $\rho_{\rm B}$ is the fraction of dividends which is not paid back by the borrower. Thus, we obtain
\begin{equation}
F^{\rm f}_t(T) = \frac{S^{\rm f}_t}{P_t(T;z^{\rm f})} - \sum_k \frac{P_t(t_k;z^{\rm f})}{P_t(T;z^{\rm f})} \left(1 - \rho_{\rm B}\right) Q^{\rm f}_t(t_k) \ind{t_k<T}
\end{equation}%

\subsection{Stock Borrowing Strategy}
Consider the opposite problem of hedging a forward contract in which at maturity you have to receive the asset in change of cash. You can borrow the asset from a third party (the lender) and sell it in the market to obtain a short position. At maturity you can use the collateral account to buy the asset and give it back to the lender. In a stock borrowing (SB) strategy, i.e. a repo contract, we need to fund the collateral and the repo fees. The borrower does not own the asset, since it is sold in the market at inception, so that it only needs to pay back the required dividend fraction to the lender. We have the same result as in the previous example, namely
\begin{equation}
q^{\rm h}_k \doteq \left(1 - \rho_{\rm B}\right) Q^{\rm h}_k
\end{equation}%
Thus, we obtain
\begin{equation}
F^{\rm f}_t(T) = \frac{S^{\rm f}_t}{P_t(T;z^{\rm f})} - \sum_k \frac{P_t(t_k;z^{\rm f})}{P_t(T;z^{\rm f})} \left(1 - \rho_{\rm B}\right) Q^{\rm f}_t(t_k) \ind{t_k<T}
\end{equation}%

\section{Equity Total Return Swaps}
\label{sec:trs}

In an equity Total Return Swap (TRS) the two parties agree in exchanging the performance of an equity asset with a Libor-indexed funding leg. If a dividend is paid before contract maturity, the equity receiver pays back a fraction $\rho_{\rm T}$ of the dividend to the equity payer. The contract is expressed in the same currency of the underlying asset, it is cash settled and it is usually collateralized. The currency of the collateral can be different from that of the contract.

We call $\{T_1,\ldots,T_n\}$ the payment dates of the equity performance, while $\{T'_1,\ldots,T'_{n'}\}$ are the payment dates of the funding leg and we assume that $T_0 = T'_0 = 0$ is the start date of the contract. There are two versions of the TRS: a first version with constant notional and a second version with constant quantity (or resetting notional).

The constant notional version of the TRS implies that the quantity of shares to be included in the hedging strategy at each coupon period should be rebalanced in order to keep the notional constant. In this case, the payoff of an equity-receiver TRS can be written by specifying the contractual coupons:
\begin{eqnarray}\label{ConstNotTRS}
\pi^{\rm TRSN,f}_t
&:=& \sum_i \left( \frac{S^{\rm f}_{T_i}}{S^{\rm f}_{T_{i-1}}} - 1 \right) \ind{t>{T_i}} + (1-\rho_{\rm T}) \sum_k \frac{\chi^{\rm h}_{t_k}}{\chi^{\rm f}_{t_k}} \frac{Q^{\rm h}_k}{S^{\rm f}_{\eta(t_k)}} \ind{t>t_k}\\\nonumber
& & - \sum_j x'_j\left( L^{\rm f}_{T'_j} + K^{\rm f} \right) \ind{t>T'_j}
\end{eqnarray}%
where $x'_j := T'_j-T'_{j-1}$ for any $j$, $L^{\rm f}_{T'_j}$ is the Libor rate fixing at $T'_{j-1}$ and paying at $T'_j$, and we define the function $\eta(T) := \max\{T_i : T_i<T\}$. For an equity-payer TRS, the payoff has opposite sign.

The constant quantity version of the TRS assumes that the quantity of shares is kept constant during the life of the contract and that the notional is accordingly reset at each coupon date according to the current price of the stock. In this case, the payoff for a quantity $1/S^{\rm f}_{0}$ of shares reads
\begin{eqnarray}\label{resetTRS}
\pi^{\rm TRSQ,f}_t
&:=& \sum_i \left( \frac{S^{\rm f}_{T_i}-S^{\rm f}_{T_{i-1}}}{S^{\rm f}_{0}}\right) \ind{t>{T_i}} + (1-\rho_{\rm T}) \sum_k \frac{\chi^{\rm h}_{t_k}}{\chi^{\rm f}_{t_k}} \frac{Q^{\rm h}_k}{S^{\rm f}_{0}} \ind{t>t_k}\\\nonumber
& & - \sum_j x'_j\left( L^{\rm f}_{T'_j} + K^{\rm f} \right) \frac{S^{\rm f}_{\eta(T'_j)}}{S^{\rm f}_0} \ind{t>T'_j}
\end{eqnarray}%

Additionally, we consider the Tobin tax, to be paid each time the hedging strategy requires we buy shares on the market. We denote the Tobin tax rate with $\tau$ and its contribution, whose specific form depends on the hedging strategies, can be included in the dividend process: 
\begin{equation}
\pi^{\rm f}_t := \pi^{\rm TRS,f}_t + \pi^{\rm Tobin,f}_t
\end{equation}%
The pricing equations of the previous section hold for any security. In particular, if we name $V^{\rm f}_t$ the TRS price in currency $\rm f$, $\beta$ the TRS collateral haircut and $c^{\rm g}_t$ the collateral accrual rate expressed in currency $\rm g$, we can compute the differential of $V^{\rm f}(t,S^{\rm f}_t)$ in the risk-neutral measure of the bank account currency, change measure to the risk-neutral measure in currency $\rm f$ and obtain
\begin{equation}
{\cal L}_{z^{\rm f}}V^{\rm f}_t = \left[-\beta_t r^{\rm f}_t + (1+\beta_t) (c^{\rm g}_t+\nu^{\rm g}_t-\nu^{\rm f}_t)\right]V^{\rm f}_t-\partial_t{\pi^{\rm f}_t}
\end{equation}%
where
\begin{equation}
{\cal L}_{z^{\rm f}} := \partial_t + z^{\rm f}S_t^{\rm f}\partial_{S} + \frac{1}{2} \partial_t \langle S^f, S^f \rangle \partial^2_{SS}
\end{equation}%
Thus, by Feynman-Ka\v{c} theorem, we get
\begin{equation}\label{eq:TRSprice}
V^{\rm f}_t = \ExT{t}{\mathbb{Q}^{\rm f}}{\int_t^T D(t,u;y^{\rm f}) \,d\pi^{\rm f}_u}
\end{equation}%
where
\begin{equation}
y^{\rm f}_t := -\beta_t r^{\rm f}_t + (1+\beta_t) c^{\rm f}_t\;, \quad c^{\rm f} = c^{\rm g}_t+\nu^{\rm g}_t-\nu^{\rm f}_t
\end{equation}%

\subsection{Constant Notional TRS}
\subsubsection{Equity Receiver TRS}
We consider an equity-receiver constant notional TRS and we select a stock borrowing startegy to hedge: at each coupon period the difference in the number of shares with respect to the number of shares of the previous period has to be bought/sold, depending on the current value of the stock. The Tobin tax contribution shows up only when this difference is positive:
\begin{equation}\label{eq:TobinConstNot}
\pi^{\rm Tobin,f}_t = -\tau \left[\sum_{i=1}^{n-1} \left( \frac{1}{S^{\rm f}_{T_{i-1}}} - \frac{1}{S^{\rm f}_{T_{i}}} \right)^+S^{\rm f}_{T_{i}}\, \ind{t>{T_i}} + \frac{S^{\rm f}_{T_{n}}}{S^{\rm f}_{T_{n-1}}}\,\ind{t>{T_n}}\right]
\end{equation}%
Substituting the expression for contractual coupons (\ref{ConstNotTRS}) and (\ref{eq:TobinConstNot}) in (\ref{eq:TRSprice}), after few manipulations we get
\begin{eqnarray}
V^{\rm f}_t
&=& \sum_i x_i P_t(T_i;y^{\rm f}) Z^{\rm f}_t(T_i) - \sum_j x'_j P_t(T'_j;y^{\rm f}) \left( L^{\rm f}_t(T'_j) + K^{\rm f} \right)\nonumber\\
& & + \sum_{i,k} P_t(t_k;y^{\rm f}) \ExT{t}{\mathbb{Q}^{\rm f}}{ \frac{Q^{\rm f}_{T_{i-1}}(t_k)}{S^{\rm f}_{T_{i-1}}} } \left( 1 - \rho_{\rm T} - (1 - \rho) \frac{P_t(T_i;y^{\rm f}-z^{\rm f})}{P_t(t_k;y^{\rm f}-z^{\rm f})} \right) \ind{T_{i-1}\le t_k<T_i}\nonumber\\
& & - \tau \sum_{i=1}^{n-1} {\cal C}_t(T_{i-1},T_i) -  \tau P_t(T_n;y^{\rm f}) \ExT{t}{\mathbb{Q}^{\rm f}}{\frac{F_{T_{n-1}}^{\rm f}(T_n)}{S^{\rm f}_{T_{n-1}}}}
\end{eqnarray}%
where we introduce the forward Libor rate $L^{\rm f}_t(T)$,
we denote with ${\cal C}_t(T_{i-1},T_i)$ the price of a forward starting ATM call on the performance which fixes at $T_{i-1}$ and pays at $T_i$, we define $\rho =\rho_I = \rho_B$ and the effective forward rate $Z^{\rm f}_t(T_i)$ as
\begin{equation}\label{ZSB}
Z^{\rm SB,f}_t(T_i) := \frac{1}{x_i} \left( \frac{P_t(T_{i-1};z^{\rm f})}{P_t(T_i;z^{\rm f})} - 1 \right)
\end{equation}%
where $x_i := T_i-T_{i-1}$ for any $i$.

The par rate for TRS contracts is then given by
\begin{equation}
K^{\rm f} = \frac{\sum_i x_i P_t(T_i;y^{\rm f}) Z^{\rm f}_t(T_i) - \sum_j x'_j P_t(T'_j;y^{\rm f}) L^{\rm f}_t(T'_j) + {\cal Q}^{\rm f}_t(t,T_n) }{\sum_j x'_j P_t(T'_j;y^{\rm f})}
\end{equation}%
where the dividend and Tobin tax funding costs can be defined as
\begin{eqnarray}
{\cal Q}^{\rm f}_t(t,T_n) &:=& \sum_{i,k} P_t(t_k;y^{\rm f}) \ExT{t}{\mathbb{Q}^{\rm f}}{ \frac{Q^{\rm f}_{T_{i-1}}(t_k)}{S^{\rm f}_{T_{i-1}}} } \left( 1 - \rho_{\rm T} - (1 - \rho) \frac{P_t(T_i;y^{\rm f}-z^{\rm f})}{P_t(t_k;y^{\rm f}-z^{\rm f})} \right) \ind{T_{i-1}\le t_k<T_i} \nonumber\\
& & - \tau \left(\sum_{i=1}^{n-1} {\cal C}_t(T_{i-1},T_i) + P_t(T_n;y^{\rm f})\ExT{t}{\mathbb{Q}^{\rm f}}{\frac{F_{T_{n-1}}^{\rm f}(T_n)}{S^{\rm f}_{T_{n-1}}}}\right)
\end{eqnarray}%

\begin{remark}
We can approximate the spots inside the expected values in (\ref{DTfunding}) with the forwards (neglecting any convexity adjustment) and, therefore, the expected dividends with $Q^{\rm f}_{t}(t_k)$. Otherwise, we can choose to rely on Black's dynamics and explicitly compute the expected values, which introduces a dependence on the volatility of the stock. In the same way, regarding the prices of the calls for the Tobin tax valuation, we can approximate them with their intrinsic value, retaining the dependence on the forward prices, otherwise we can use Black formula. The following numerical examples will make use of Black dynamics. 
\end{remark}

\subsubsection{Equity Payer TRS}
In the previous derivation the TRS is equity receiver. In the case of an equity-payer TRS we can adopt the buy-and-hold or stock-lending hedging strategy. In this case, the Tobin tax contribution changes as
\begin{equation}
\pi^{\rm Tobin,f}_t = -\tau \left[1+\sum_{i=1}^{n-1} \left( \frac{1}{S^{\rm f}_{T_i}} - \frac{1}{S^{\rm f}_{T_{i-1}}} \right)^+S^{\rm f}_{T_{i}}\,\ind{t>{T_i}}\right]
\end{equation}%
giving rise to a sequence of forward starting puts (instead of calls) on the performance.
Moreover, if we adopt a buy-and-hold strategy, we get the following effective forward rate:
\begin{equation}\label{ZBH}
Z^{\rm BH,f}_t(T_i) := R^{\rm f}_t(T_i) := \frac{1}{x_i} \left( \frac{P_t(T_{i-1};r^{\rm f})}{P_t(T_i;r^{\rm f})} - 1 \right)
\end{equation}%
otherwise, if we adopt a stock-lending strategy, we get
\begin{equation}\label{ZSL}
Z^{\rm SL,f}_t(T_i) := \frac{1}{x_i} \left( \frac{P_t(T_{i-1};z^{\rm f})}{P_t(T_i;z^{\rm f})} - 1 \right)
\end{equation}%
It is also possible to make a blending of the two strategies, where the blended rates and dividend taxes are given by convex combinations of the BH and SL ones with a weight $w$:
\begin{eqnarray}%
z^{\rm f}_t &=& w \left[-\alpha_t r^{\rm f}_t + (1+\alpha_t) c^{\rm f}_t - \ell^{\rm f}_t\right] + (1-w) r^{\rm f}_t\\
\rho &=& w \rho_B+(1-w)\rho_I
\end{eqnarray}

\subsection{Resetting Notional TRS}
If we consider a resetting notional equity-receiver TRS (\ref{resetTRS}), the time $t$ price is given by
\begin{eqnarray}
V^{\rm f}_t &=& \sum_i P_t(T_i;y^{\rm f}) \frac{F^{\rm f}_t(T_i)-F^{\rm f}_t(T_{i-1})}{S^{\rm f}_0} +  (1-\rho_{\rm T}) \sum_k P_t(t_k;y^{\rm f}) \frac{Q^{\rm f}_t(t_k)}{S^{\rm f}_0}  \ind{t_k<T_n} \nonumber\\
& & - \sum_j x'_j P_t(T'_j;y^{\rm f}) \left( L^{\rm f}_t(T'_j) + K^{\rm f} \right)\frac{F^{\rm f}_t(\eta(T'_j))}{S^{\rm f}_0} -  \tau P_t(T_n;y^{\rm f}) \frac{F^{\rm f}_t(T_n)}{S^{\rm f}_0}
\end{eqnarray}%
where we used the fact that the Tobin tax is now given by
\begin{equation}
\pi^{\rm Tobin,f}_t = -\tau\, \frac{S^{\rm f}_{T_n}}{S^{\rm f}_{0}}\,\ind{t>T_n}
\end{equation}%
since no rebalancing of the number of shares happens during the life of the TRS.
Similar computations as in the case of constant notional TRS lead to the following result for the par spread:
\begin{equation}
K^{\rm f} = \frac{\sum_i x_i P_t(T_{i-1};y^{\rm f}-z^{\rm f}) P_{T_{i-1}}(T_i;y^{\rm f}) Z^{\rm f}_t(T_i) - \sum_j x'_j P_t(T'_j;y^{\rm f}) L^{\rm f}_t(T'_j)\frac{F^{\rm f}_t(\eta(T'_j))}{S^{\rm f}_0} + {\cal Q}^{\rm f}_t(t,T_n) }{\sum_j x'_j P_t(T'_j;y^{\rm f})\frac{F^{\rm f}_t(\eta(T'_j))}{S^{\rm f}_0}}
\end{equation}%
where the dividend and Tobin tax funding costs are now defined as
\begin{eqnarray}
{\cal Q}^{\rm f}_t(t,T_n) &:=& \sum_{i,k} P_t(t_k;y^{\rm f}) \frac{Q^{\rm f}_t(t_k)}{S^{\rm f}_0} \left( 1 - \rho_{\rm T} - (1 - \rho) \frac{P_t(T_i;y^{\rm f}-z^{\rm f})}{P_t(t_k;y^{\rm f}-z^{\rm f})} \right) \ind{T_{i-1}\le t_k<T_i} \\
& & -\sum_{i,k}x_i P_t(T_i;y^{\rm f}) Z_t^{\rm f}(T_i) (1-\rho) \frac{Q^{\rm f}_t(t_k)}{S^{\rm f}_0} \frac{P_t(t_k;z^{\rm f})}{P_t(T_{i-1};z^{\rm f})} \ind{t_k<T_{i-1}} \nonumber\\
& & - \tau \left(P_t(T_n;y^{\rm f}-z^{\rm f}) + \sum_{k} P_t(t_k;z^{\rm f})P_t(T_n;y^{\rm f}-z^{\rm f}) (1-\rho) \frac{Q^{\rm f}_t(t_k)}{S^{\rm f}_0}\ind{t_k<T_n}\right)\nonumber
\end{eqnarray}%
In particular, if the equity and funding legs share the same schedule, i.e. $x = x'$, $T_i = T'_j$ and $\eta(T'_j) = T_{i-1}$, the above formula simplifies as:
\begin{equation}\label{spread}
K^{\rm f} = \frac{\sum_i x_i P_t(T_{i-1};y^{\rm f}-z^{\rm f}) P_{T_{i-1}}(T_i;y^{\rm f}) \left(Z^{\rm f}_t(T_i)-L^{\rm f}_t(T_i)\right) + {\cal Q}^{\rm f}_t(t,T_n) }{\sum_i x_i P_t(T_i;y^{\rm f})\frac{F^{\rm f}_t(T_{i-1})}{S^{\rm f}_0}}
\end{equation}%
where
\begin{eqnarray}\label{DTfunding}
{\cal Q}^{\rm f}_t(t,T_n) &:=& \sum_{i,k} P_t(t_k;y^{\rm f}) \frac{Q^{\rm f}_t(t_k)}{S^{\rm f}_0} \left( 1 - \rho_{\rm T} - (1 - \rho) \frac{P_t(T_i;y^{\rm f}-z^{\rm f})}{P_t(t_k;y^{\rm f}-z^{\rm f})} \right) \ind{T_{i-1}\le t_k<T_i} \\
& & -\sum_{i,k}x_i P_t(T_i;y^{\rm f}) \left(Z_t^{\rm f}(T_i)-L_t^{\rm f}(T_i)\right) (1-\rho) \frac{Q^{\rm f}_t(t_k)}{S^{\rm f}_0} \frac{P_t(t_k;z^{\rm f})}{P_t(T_{i-1};z^{\rm f})} \ind{t_k<T_{i-1}} \nonumber\\
& & - \tau \left(P_t(T_n;y^{\rm f}-z^{\rm f}) + \sum_{k} P_t(t_k;z^{\rm f})P_t(T_n;y^{\rm f}-z^{\rm f}) (1-\rho) \frac{Q^{\rm f}_t(t_k)}{S^{\rm f}_0}\ind{t_k<T_n}\right)\nonumber
\end{eqnarray}%
In the case of an equity-payer TRS, the same remarks as for the constant notional case hold, with the difference that the Tobin tax contribution is simply given by $\pi^{\rm Tobin,f}_t = -\tau$.

\subsection{Par Spread Expansion}
We now derive an approximated par rate to understand the funding effects. Here, we consider constant haircuts and we focus on a constant notional TRS, just to fix the notation. According to the different hedging strategies we get
\begin{equation}
Z^{\rm BH,f}_t(T_i) = \Delta R^{\rm f}_t(T_i) + E^{\rm f}_t(T_i)
\end{equation}%
and
\begin{equation}
Z^{\rm SL,f}_t(T_i) \approx -\alpha \Delta R^{\rm f}_t(T_i) + (1+\alpha) \Delta C^{\rm f}_t(T_i) + E^{\rm f}_t(T_i) - \Delta M^{\rm f}_t(T_i)
\end{equation}%
where $E$ is the one-period OIS rate, $\Delta R$ is the one-period funding spread over OIS,  $\Delta C$ is the one-period collateral spread over OIS and $\Delta M$ is the one-period repo spread over OIS. If we assume that all rates are small, and the fractions $\rho$ and $\tau$ as well, discarding all second order terms we get
\begin{eqnarray}
{\cal Q}_t(t,T_n) &\approx& \sum_{i,k} P_t(t_k;e) \ExT{t}{\mathbb{Q}^{\rm f}}{ \frac{Q^{\rm f}_{T_{i-1}}(t_k)}{S_{T_{i-1}}} } \gamma_t(t_k,T_i) \ind{T_{i-1}\le t_k<T_i}\nonumber\\
& & -\,\tau \text{ order terms}
\end{eqnarray}%
where $e$ is the OIS short rate and we define, according to the hedging strategy, the dividend tax impact rate as given by
\begin{equation}
\gamma^{\rm BH}_t(t_k,T_i) := \rho - \rho_{\rm T} - (1+\beta) [\Delta R^{\rm f}_t(T_i)-\Delta C^{\rm f}_t(T_i)](T_i-t_k)
\end{equation}%
and
\begin{equation}
\gamma^{\rm SL}_t(t_k,T_i) := \rho_{\rm B} - \rho_{\rm T} - \left[(\beta-\alpha)\left(\Delta R^{\rm f}_t(T_i) - \Delta C^{\rm f}_t(T_i)\right) - \Delta M^{\rm f}_t(T_i)\right](T_i-t_k)
\end{equation}%
Hence, if we have the same schedules in both legs, we can write according to the hedging strategies
\begin{equation}\label{fundeffBH}
K^{\rm f,BH} = \frac{ \sum_i x_i P_t(T_i;e) \left[ \Delta R^{\rm f}_t(T_i) - \Delta L^{\rm f}_t(T_i) \right] + {\cal Q}_t(t,T_n;\gamma^{\rm BH}) }{ \sum_i x_i P_t(T_i;e)}
\end{equation}%
and
\begin{equation}\label{fundeffSL}
K^{\rm f,SL} = \frac{ \sum_i x_i P_t(T_i;e) \left[-\alpha \Delta R^{\rm f}_t(T_i) + (1+\alpha)\Delta C_t^{\rm f}(T_i)- \Delta L^{\rm f}_t(T_i) - \Delta M^{\rm f}_t(T_i) \right] + {\cal Q}_t(t,T_n;\gamma^{\rm SL}) }{ \sum_i x_i P_t(T_i;e)}
\end{equation}%
where $\Delta L$ is the one-period Libor spread over OIS.

\section{Numerical Investigations}
\label{sec:numeric}
We now consider a concrete example of a Total Return Equity Swap and discuss the impact of some parameters and market scenarios, both related to the contract itself and to its hedge, on the par spread.

In the following, we assume a performance-payer TRS contract on BASF SE against EURIBOR 1M, with one-year maturity and monthly resetting notional. The underlying spot price is 73 EUR and a dividend payment of 3.2 EUR is scheduled during the first period of the swap. The parameters given in Table \ref{tab:param} are kept fixed during all numerical computations, while the repo dividend tax $\rho_B$, the blending parameter $w$, the repo fees $\ell$ and the levels of collateral and funding rates $c_t$ and $r_t$ are let vary. The computation is performed as of 18 April 2019.\\
\begin{table}[h]
\centering
\begin{tabular}{|l|c|}
  \hline
  TRS dividend tax & $\rho_T = 0$\\
  Investor taxation & $\rho_I = 15\%$\\
  Collateral rate & EONIA\\
  Repo haircut & $\alpha = 5\%$\\
  TRS haircut & $\beta = 0$\\
  Tobin tax & $\tau = 10\,bps$\\
  \hline
\end{tabular}
\caption{TRS parameters}\label{tab:param}
\end{table}

\begin{figure}
	\centering
	\begin{tikzpicture}
		\begin{axis}[
			xlabel=$\rho_B$ ($\%$),
			ylabel=$K$ ($\%$),
            ymin=-1.2, ymax=1.2,
            legend pos=south east]
		\addplot file {Blending1.dat};  \addlegendentry{$w = 1$};
		\addplot file {Blending08.dat}; \addlegendentry{$w = 0.8$};
        \addplot file {Blending06.dat}; \addlegendentry{$w = 0.6$};
        \addplot file {Blending04.dat}; \addlegendentry{$w = 0.4$};
        \addplot file {Blending02.dat}; \addlegendentry{$w = 0.2$};
        \addplot file {Blending0.dat};  \addlegendentry{$w = 0$};
		\end{axis}
	\end{tikzpicture}
	\caption{TRS spread $K$ vs Repo dividend tax $\rho_B$ for different values of the blending parameter $w$.}\label{fig:VaryingRepoTax}
\end{figure}
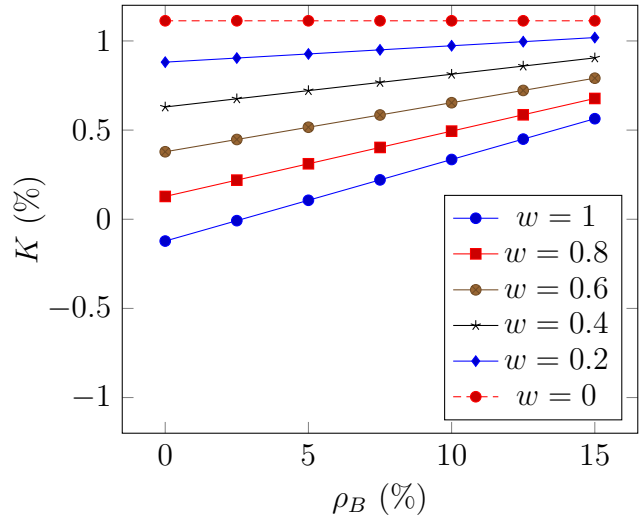

In Figure \ref{fig:VaryingRepoTax} we show how the par spread $K$ changes with the repo dividend tax $\rho_B$: each line corresponds to a specific value of $w$, \emph{i.e.} to a given mix of BH ($w=0$) and SL ($w=1$) hedging strategies. We observe that the buy-and-hold strategy achieves the highest value of $K$ for typical levels of $\rho_B$ (between 0 and $15\%$), while $K$ increases with $\rho_B$ in the case of the stock lending strategy. Indeed, from (\ref{spread}) and (\ref{DTfunding}) it is clear that the TRS spread is sensitive to the differential between taxes $\rho$ and $\rho_T$ (recall that $\rho = w\rho_B + (1-w)\rho_I$ and we are assuming $\rho_T = 0$ here).

\begin{figure}
	\centering
	\begin{tikzpicture}
		\begin{axis}[
			xlabel=$w$,
			ylabel=$K$ ($\%$),
            ymin=-1.6, ymax=1.5,
            legend pos=south west]
		\addplot file {RepoTax0.dat};  \addlegendentry{$\rho_B = 0\%$};
		\addplot file {RepoTax5.dat};  \addlegendentry{$\rho_B = 5\%$};
        \addplot file {RepoTax10.dat}; \addlegendentry{$\rho_B = 10\%$};
        \addplot file {RepoTax15.dat}; \addlegendentry{$\rho_B = 15\%$};
        \addplot file {RepoTax20.dat}; \addlegendentry{$\rho_B = 20\%$};
        \addplot file {RepoTax25.dat}; \addlegendentry{$\rho_B = 25\%$};
        \addplot file {RepoTax30.dat}; \addlegendentry{$\rho_B = 30\%$};
		\end{axis}
	\end{tikzpicture}
	\caption{TRS spread $K$ vs blending parameter $w$ for different values of the Repo dividend tax $\rho_B$.}\label{fig:VaryingBlending}
\end{figure}
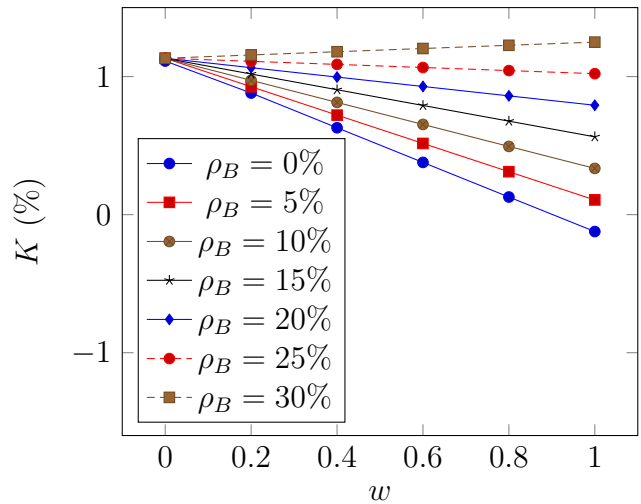

In Figure \ref{fig:VaryingBlending} we show how the par spread $K$ changes with the blending parameter $w$: each line corresponds to a specific value of $\rho_B$. We observe that, with a stock lending strategy, different values of $\rho_B$ lead to a range of variation of $K$ higher than 100 bps. However, a level of $K$ comparable or higher than that achieved by a buy and hold strategy is only possible with extreme values of $\rho_B$, so that in practice $K$ achieved with SL is always below that achieved with BH. In order to obtain a reversed behaviour, we have to change other parameters, e.g. we would need $\rho_I \leq \rho_B$.

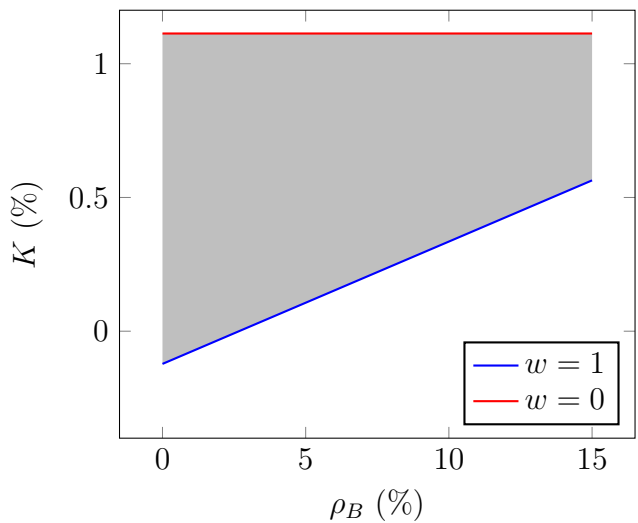
\begin{figure}
	\centering
	\begin{tikzpicture}
		\begin{axis}[
			xlabel=$\rho_B$ ($\%$),
			ylabel=$K$ ($\%$),
            ymin=-0.4, ymax=1.2,
            thick,
            legend pos=south east]
		\addplot[name path=A,blue,thick,no marks] file {Blending1.dat};  \addlegendentry{$w = 1$};
        \addplot[name path=B,red,thick,no marks] file {Blending0.dat};  \addlegendentry{$w = 0$};
		\addplot[gray!50] fill between[of=A and B];
        \end{axis}
	\end{tikzpicture}
	\caption{Accessibility region (in gray) for the TRS spread $K$ when the Repo dividend tax $\rho_B$ and the blending parameter $w$ are let vary.}\label{fig:TRSspread}
\end{figure}

The above analysis shows that the parameters $\rho_B$ and $w$ define a region of accessibility for the par spread $K$ of the TRS contract, whose values depend on the characteristics of the hedging strategy (see Figure \ref{fig:TRSspread}): suppose that $\rho_B$ is given, then a target price can be achieved with a specific choice of $w$, provided that we stay inside the gray region. The choice of $K$ can be motivated by different arguments, such as maximizing the PL or obtaining a competitive price for the client.
 
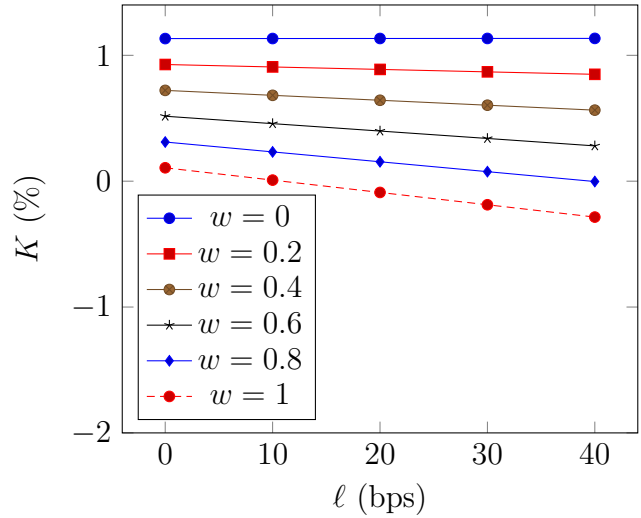
\begin{figure}
	\centering
	\begin{tikzpicture}
		\begin{axis}[
			xlabel=$\ell$ (bps), 
			ylabel=$K$ ($\%$),
            ymin=-2, ymax=1.4,
            legend pos=south west]
		\addplot file {RepoFee0.dat};  \addlegendentry{$w = 0$};
		\addplot file {RepoFee02.dat}; \addlegendentry{$w = 0.2$};
        \addplot file {RepoFee04.dat}; \addlegendentry{$w = 0.4$};
        \addplot file {RepoFee06.dat}; \addlegendentry{$w = 0.6$};
        \addplot file {RepoFee08.dat}; \addlegendentry{$w = 0.8$};
        \addplot file {RepoFee1.dat};  \addlegendentry{$w = 1$};
		\end{axis}
	\end{tikzpicture}
	\caption{TRS spread $K$ vs Repo fees $\ell$ for different values of the blending parameter $w$. The Repo dividend tax is set to $5\%$.}\label{fig:VaryingRepoFees}
\end{figure}

In Figure \ref{fig:VaryingRepoFees} we show how the par spread $K$ decreases with increasing repo fees. We show the results for the case $\rho_B = 5\%$.

\begin{figure}
	\centering
	\begin{tikzpicture}
		\begin{axis}[
			xlabel=$w$,
			ylabel=$K$ ($\%$),
            ymin=-0.2, ymax=1.4,
            thick,
            legend pos=south west]
		\addplot[name path=A,blue,thick,no marks] file {CreditUp.dat};  \addlegendentry{$r_t + 10bps$};
        \addplot[name path=B,red,thick,no marks] file {CreditDown.dat};  \addlegendentry{$r_t - 10bps$};
		\addplot[gray!50] fill between[of=A and B];
        \end{axis}
	\end{tikzpicture}
	\caption{TRS spread $K$ in terms of blending parameter $w$ for different levels of the funding rates. The TRS is fully collateralized. The Repo dividend tax is set to $5\%$}\label{fig:VaryingFunding}
\end{figure}
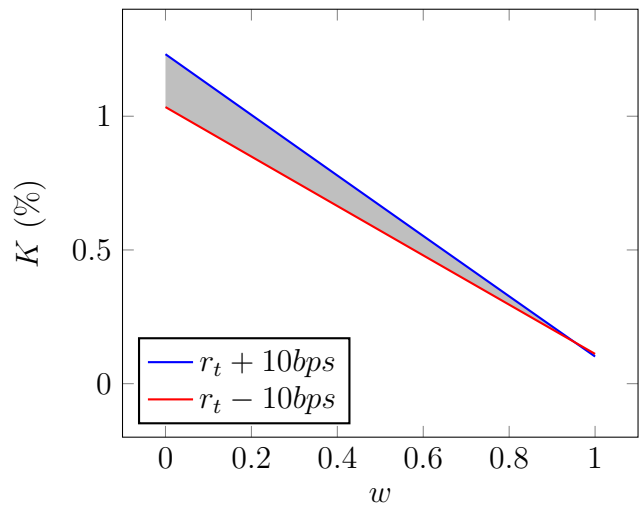

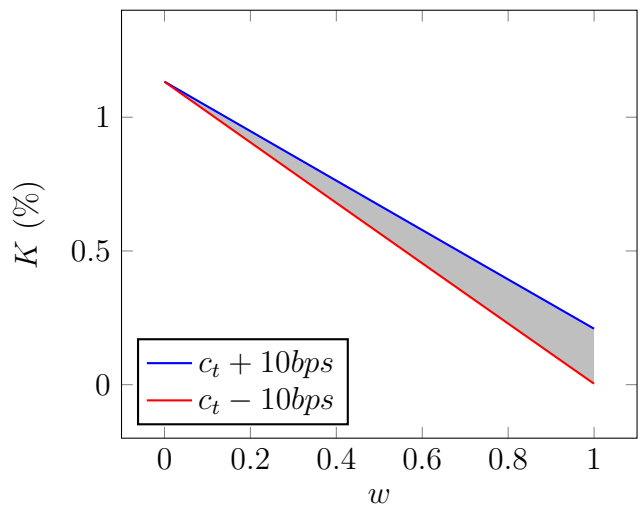
\begin{figure}
	\centering
	\begin{tikzpicture}
		\begin{axis}[
			xlabel=$w$,
			ylabel=$K$ ($\%$),
            ymin=-0.2, ymax=1.4,
            thick,
            legend pos=south west]
		\addplot[name path=A,blue,thick,no marks] file {CollUp.dat};  \addlegendentry{$c_t + 10bps$};
        \addplot[name path=B,red,thick,no marks] file {CollDown.dat};  \addlegendentry{$c_t - 10bps$};
		\addplot[gray!50] fill between[of=A and B];
        \end{axis}
	\end{tikzpicture}
	\caption{TRS spread $K$ in terms of blending parameter $w$ for different levels of the collateral rate. The TRS is fully collateralized. The Repo dividend tax is set to $5\%$}\label{fig:VaryingCollateral}
\end{figure}

In Figures \ref{fig:VaryingFunding} and \ref{fig:VaryingCollateral} we show how the par spread $K$ changes with different levels of the funding and collateral rates: of course, the entity of this impact depends on the degree of collateralization and on the value of the blending between SL and BH strategies: in the present case, even though the TRS contract is perfectly collateralized, there is still an impact of the funding rate $r_t$ due to the funding mechanism of the buy-and-hold and of the (over-collateralized) stock lending strategy. Increasing funding and collateral rates imply a proportional increase in the value of the TRS par spread. We again show the results for the case $\rho_B = 5\%$.

All the results are coherent with the behaviour derived in (\ref{fundeffBH}) and (\ref{fundeffSL}).

The equity sensitivities of the par spread are much smaller than those of the rate curves (few basis points for $10\%$ changes in the spot and dividend levels). 

Finally, we notice that, even though we described here the case of a resetting notional TRS, similar results hold for the constant-notional version.

\section{Conclusions and Further Developments}
\label{sec:conclusion}
In this work we have discussed the impact of funding costs on linear equity derivatives such as Total Return Swaps. We adopted a martingale pricing approach in the presence of dividend-paying assets and collateralized contracts, including partial collateralization of the derivative and its hedge (the latter assumed to be performed through repo transactions). We have shown that funding costs are reflected in the choice of the discounting curves: in the case of partially collateralized TRS, standard arguments imply that a correction proportional to the funding spread of the bank is added to the OIS-discounted TRS spread. Moreover, the funding costs of the hedge also affect the TRS spread, not only through the funding spreads coming from the repo-adjusted blended rate (\ref{eq:reporate}), but also through the differential on dividend taxes and through the Tobin tax, as it is clear from equations (\ref{fundeffBH}) and (\ref{fundeffSL}). The choice of the hedging strategy, together with the differences between the involved tax regimes, allow for different values of the TRS spread, hence of its price. These arguments are important when a profitability analysis is carried out before a trade is closed.

In the present analysis, we have neglected any contribution coming from counterparty risk: while it is justified on the derivative side, since TRS are usually traded under CSA, a residual CVA is still present on the hedge side in the case of stock lending/borrowing, since the latter transactions are usually over-collateralized. We plan to address the problem of pricing the default of the counterparty in the stock lending transaction in an updated version of this paper.

\bibliographystyle{nonumber}

\end{document}